\documentclass[referee,a4paper,12pt,traditabstract]{swsc} 

\usepackage{graphicx}
\usepackage{txfonts}
\usepackage{subfigure}
\usepackage{epstopdf}
\usepackage[authoryear,round]{natbib}
\usepackage[backref]{hyperref}
\usepackage{url}
\usepackage{lscape}
\usepackage{afterpage}

\hypersetup{colorlinks=true,citecolor=cyan,urlcolor=cyan,linkcolor=blue}

\begin{document}


   \title{Verification of real-time WSA--ENLIL+Cone simulations of CME arrival-time at the CCMC from 2010--2016}
   
   \titlerunning{CME arrival-time validation}

   \authorrunning{Wold}

   \author{Alexandra M. Wold
          \inst{1,2}
          M. Leila Mays\inst{3}
          A. Taktakishvili\inst{4,3}
          L. K. Jian\inst{5,3}
          D. Odstrcil\inst{3,6}
          P. MacNeice\inst{3}
          }

   \institute{American University, Physics Department
         \and
             University of Colorado Boulder, Aerospace Engineering Sciences\\
              \email{\href{mailto:alexandra.wold@colorado.edu}{alexandra.wold@colorado.edu}}\thanks{Begining in August 2017.}
         \and
             NASA Goddard Space Flight Center
         \and
             Catholic University of America
         \and
             University of Maryland College Park
         \and
             George Mason University
             }


 
  \abstract
       {The Wang-Sheeley-Arge (WSA)-ENLIL+Cone model is used extensively in space weather operations world-wide to model CME propagation. As such, it is important to assess its performance. We present validation results of the WSA--ENLIL+Cone model installed at the Community Coordinated Modeling Center (CCMC) and executed in real-time by the CCMC space weather team. CCMC uses the WSA--ENLIL+Cone model to predict CME arrivals at NASA missions throughout the inner heliosphere.  In this work we compare model predicted CME arrival-times to in-situ ICME leading edge measurements at {\it STEREO-A}, {\it STEREO-B}, and Earth ({\it Wind} and {\it ACE}) for simulations completed between March 2010--December 2016 (over 1,800 CMEs).  We report hit, miss, false alarm, and correct rejection statistics for all three locations. For all predicted CME arrivals, the hit rate is 0.5, and the false alarm rate is 0.1. For the 273 events where the CME was predicted to arrive at Earth, STEREO-A, or STEREO-B, and was actually observed (hit event), the mean absolute arrival-time prediction error was $10.4 \pm 0.9$ hours, with a tendency to early prediction error of -4.0 hours. We show the dependence of the arrival-time error on CME input parameters. We also explore the impact of the multi-spacecraft observations used to initialize the model CME inputs by comparing model verification results before and after the {\it STEREO-B} communication loss (since September 2014) and {\it STEREO-A} sidelobe operations (August 2014-December 2015). There is an increase of 1.7 hours in the CME arrival time error during single, or limited two-viewpoint periods, compared to the three-spacecraft viewpoint period. This trend would apply to a future space weather mission at L5 or L4 as another coronagraph viewpoint to reduce CME arrival time errors compared to a single L1 viewpoint.}

   \keywords{MHD --
                Modeling --
                Validation --
                Forecasting --
                Coronal Mass Ejection (CME)
               }

   \maketitle

\section{Introduction}
\label{sec:intro}

The Wang--Sheeley--Arge (WSA) coronal model \citep{arge2000,arge2004} coupled with the global heliospheric ENLIL solar-wind model \citep{ods1996,odstrcil1999_1,odstrcil1999_2,odstrcil2003,odstrcil2004} has been used extensively in space weather operations world-wide. Space weather models provide forecast capabilities that greatly enhance satellite and ground-based observations. It is essential for both model users and developers to understand the limitations and capabilities of these models. In order to measure model performance, generally the model output is compared to a measurable parameter and skill scores are computed. Results from model verification are also helpful as feedback to the model developers, setting benchmarks for the current state of a model, and determining the usefulness and capabilities of a model for operations. 

  Previous studies have also assessed aspects of WSA--ENLIL+Cone model performance, typically with a smaller sample size of selected events, and in a non-real-time setting. \cite{taktakishvili2009} studied the performance of the ENLIL+Cone model in modeling the propagation of coronal mass ejections (CMEs) in the heliosphere by comparing the results of the simulation with $ACE$ satellite observations. They evaluated the results of the ENLIL+Cone model for 14 fast CME events and found more earlier arrival predictions (9 out of 14) than late arrival predictions. The errors on the earlier arrivals were, on the average, larger than those of the late arrival predictions. The average absolute error was approximately 6 hours for the total set.  \cite{millward2014} assessed 25 CME events during the first year of WSA--ENLIL operations at NOAA Space Weather Prediction Center (October 2011--October 2012) and found an average error of 7.5 hours. \cite{mays2015} assesses the WSA--ENLIL+Cone ensemble modeling of CMEs in the real-time CCMC setting. The ensemble modeling method provides a probabilistic forecast of CME arrival time and an estimation of arrival-time uncertainty from the spread and distribution of predictions, as well as forecast confidence in the likelihood of CME arrival. For 17 predicted CME arrivals, the mean absolute arrival-time prediction error was 12.3 hours with an early bias of -5.8 hours. \cite{vrsnak2014} compares a drag-based model (DBM) to the WSA-ENLIL+Cone model, and shows that the average arrival time error for DBM is about 14 hrs. They also show that DBM performs similarly to ENLIL during low solar activity periods, but ENLIL performs better as solar activity increases.

In interpreting the results of arrival time error assessment for the WSA--ENLIL+Cone model, it is important to examine all of the factors that can contribute to this uncertainty. \cite{mays2015} found that the reliability of ensemble CME-arrival predictions was heavily dependent on the initial distribution of CME input parameters, especially speed and width. \cite{millward2014} provides an analysis of the impact that CME input parameters have on ENLIL’s arrival time predictions, as well as the importance of multi-viewpoint coronagraph imagery in determining these parameters. \cite{millward2014} found that when CMEs are measured from a single viewpoint with cone analysis, it is difficult to establish objectively the correct ellipse that should be applied to a given halo CME. Any uncertainties in the projected elliptical face of the cone were seen to lead to large errors in calculations of cone angle and radial velocity. \cite{millward2014} reports that three viewpoint measuring of CMEs improves the accuracy of measurements, and that for outputs of models like ENLIL to be meaningful, the accuracy of key CME parameters is essential. 

Beyond the improvement in arrival time prediction expected when using three coronagraph viewpoints instead of one or two, tracking CME propagation beyond coronagraph field of views leads to improved arrival time predictions. \cite{moestl2014} found that predicting CME speeds and arrival times with heliospheric images gives more accurate results than using projected initial speeds from coronagraph measurements on the order of 12 hours for the arrival times. By comparing predictions of speed and arrival time for 22 CMEs to the corresponding interplanetary coronal mass ejection (ICME) measurements at in-situ observatories, they found the absolute difference between predicted and observed ICME arrival times was $8.1 \pm 6.3$ hours (RMS = 10.9 hrs), with their empirical corrections improving their performance for the arrival times to $6.1 \pm 5.0$ hours (root mean square error RMSE = 7.9 hrs). While error in arrival time predictions decreases with this heliospheric image tracking, the prediction lead time decreases. \cite{colaninno2013} used a variety of non-real-time methods to evaluate CME arrival-time predictions based on a linear fit above a height of 50 solar radii (R$_{\odot}$) to multi-viewpoint imaging data analysis only, and found an average absolute error of 6 hrs for seven out of nine CMEs, and 13 hrs for the full sample of nine CMEs. 

\cite{moestl2014} and \cite{millward2014} mention important factors other than CME parameter inputs that affect prediction quality. The interaction of multiple CMEs  \citep{lee2013}, the lack of ejecta magnetic structure, differences between a direct hit or a glancing blow \citep{moestl2015,mays2015b}, and prediction errors from other model limitations should be considered. Reliable characterization of the ambient solar wind flow is also necessary for simulating transients and CME propagation \citep{lee2013,mays2015}. 

In this article we evaluate the performance of the WSA--ENLIL+Cone model installed at the Community Coordinated Modeling Center (CCMC) and executed in real-time by the CCMC space weather team from March 2010--December 2016.  The CCMC, located at NASA Goddard Space Flight Center, is an interagency partnership to facilitate community research and accelerate implementation of progress in research into space weather operations. The CCMC space weather team is a CCMC sub-team that provides space weather services to NASA robotic mission operators and science campaigns and prototypes new models, forecasting techniques, and procedures. The CCMC space weather team began performing real-time WSA--ENLIL+Cone simulations in March 2010, marking the start of our verification period. The CCMC also serves the {\it CME Scoreboard} website (\href{http://kauai.ccmc.gsfc.nasa.gov/CMEscoreboard}{{\sf kauai.ccmc.gsfc.nasa.gov/CMEscoreboard}}) to the research community who may submit CME arrival-time predictions in real-time for different forecasting methods.

In Section \ref{sec:model} we provide a brief description of the WSA--ENLIL+Cone model simulations and input parameters. We describe the methodology of our verification study in Section \ref{sec:method}. We compute the average error (bias) and the average absolute arrival time error in Section \ref{sec:error} and examine the dependence of these errors on CME input parameters, including direction, speed, and width. In Section \ref{sec:subsets} we determine significance of multi-spacecraft observations for deriving CME parameters used to initialize the model.  This is done by comparing arrival time errors before and after the  {\it Solar TErrestrial RElations Observatory} \citep[{\it STEREO:}][]{stereo} 
{\it Behind (B)} communication loss in September 2014, and during {\it STEREO-Ahead (A)} sidelobe operations, from August 2014--December 2015.  In Section \ref{sec:skillscores}, we report CME arrival hit, miss, false alarm, and correct rejection statistics and skill scores.   For Earth arrivals, verification of $K_{\rm P}$ forecasts are presented in Section \ref{sec:kp}.  Finally, in Section \ref{sec:conclusions} we summarize our results.

\section{WSA--ENLIL+Cone Model Simulations}
\label{sec:model}
The WSA--ENLIL+Cone model consists of two parts, the  Wang-Sheeley-Arge (WSA) coronal model that approximates solar wind outflow at 21.5 R$_{\odot}$, beyond the solar wind critical point, and the ENLIL 3D MHD numerical model that provides a time-dependent description of the background solar wind plasma and magnetic field into which a CME can be inserted at the inner boundary of 21.5 R$_{\odot}$. A common method to estimate the 3D CME kinematic and geometric parameters is to assume that the geometrical CME properties are approximated by the Cone model \citep{zhao2002,xie2004}, which assumes isotropic expansion, radial propagation, and constant CME cone angular width. Generally, a CME disturbance is inserted in the WSA--ENLIL model as slices of a homogeneous spherical plasma cloud with uniform velocity, density, and temperature as a time-dependent inner boundary condition  with a steady magnetic field. Three dimensional CME parameters were determined using the Stereoscopic CME Analysis Tool (StereoCAT) \citep{mays2015} and the NOAA Space Weather Prediction Center CME Analysis Tool (CAT) \citep{millward2014}. For most of the simulations in this study, ENLIL model version 2.7 was used with ambient settings ``a4b1", together with WSA version 2.2.  A small subset of simulations prior to May 2011 (101 simulations; 34 hits) were performed with an earlier ENLIL version 2.6 using the ambient setting of ``a3b2". \cite{jian2011} describes the model version differences in more detail.

In this study, the results of the simulations are compared to in-situ ICME arrivals near Earth, {\it STEREO-A} and {\it STEREO-B} from March 2010 through December 2016.  This set includes simulations of about 1,800 CMEs.  Coronagraph observations from the {\it SOlar and Heliospheric Observatory} \citep[{\it SOHO:}][]{soho} spacecraft at L1 ahead of Earth, and also the {\it STEREO-A} and {\it B} spacecraft trailing ahead and behind Earth's orbit were used.  Since these simulations are all made in real-time by space weather forecasters, many slow CMEs (under 500 km\,s$^{-1}$) and CMEs out of the ecliptic plane (narrow CMEs at latitudes $>$ 25$^{\circ}$) may not be modeled.  Additionally, all of the coronagraph derived CME measurements were derived in real-time, often inferred from just a few data points due to real-time data gaps. Finally, there are many different space weather forecasters with varying levels of experience producing the measurements that were used to initialize the simulations.

As described in \cite{emmons2013} and \cite{mays2015}, for Earth-directed CMEs, WSA--ENLIL+Cone model outputs are used to compute an estimate of the geomagnetic $K_{\rm P}$ index using the \cite{newell2007} coupling function. Three magnetic field clock-angle scenarios of 90$^{\circ}$ (westward), 135$^{\circ}$ (southwestward), and 180$^{\circ}$ (southward) are calculated to compute the $K_{\rm P}$. Verification results of the $K_{\rm P}$ predictions are presented in Section \ref{sec:kp}.

\section{Verification Methodology}
\label{sec:method}
The quality of model performance is evaluated by comparing the model output to the observed ICME arrival time. Both the predicted and observed arrival times refer to the arrival of the leading edge of the ICME shock or compression wave and not the magnetic cloud start time (if it exists). The CCMC's publicly available {\it Space Weather Database Of Notifications, Knowledge, Information} {\it (DONKI)} (\href{https://ccmc.gsfc.nasa.gov/donki/}{{\sf ccmc.gsfc.nasa.gov/donki}}) is populated by the CCMC space weather team and contains space weather relevant flares,  SEPs (solar energetic particles), CME parameters, WSA-ENLIL+Cone CME simulations, ICME arrivals, high speed streams, modeled magnetopause crossings, and radiation belt enhancements. All simulations used in this study were conducted in real--time and recorded in DONKI, while some of the observed ICME arrivals at L1, {\it STEREO-A}, and {\it STEREO-B} were added post-event. In order to determine observed ICME arrivals that may have been missing or incorrect in DONKI, existing ICME catalogs were compared to those already documented in the DONKI database. These included the \cite{jian2006,jian2011} $Wind/ACE$ and STEREO ICME catalogs \citep{jian2013} (\href{http://www-ssc.igpp.ucla.edu/forms/stereo/stereo_level_3.html}{{\sf www-ssc.igpp.ucla.edu/forms/stereo/stereo\_level\_3.html}}, \href{ftp://stereodata.nascom.nasa.gov/pub/ins_data/impact/level3/README.html}{{\sf ftp://stereodata.nascom.nasa.gov/pub/ins\_data/impact/level3/README.html}}), \cite{richardson2010} ICME catalog (\href{http://www.srl.caltech.edu/ACE/ASC/DATA/level3/icmetable2.htm}{{\sf www.srl.caltech.edu/ACE/ASC/DATA/level3/icmetable2.htm}}), International Study of Earth-Affecting Solar Transients (ISEST) catalog (\href{http://solar.gmu.edu/heliophysics/index.php/The_ISEST_Master_CME_List}{{\sf solar.gmu.edu/heliophysics/index.php/The\_ISEST\_Master\_CME\_List}}), and the \cite{nieves2016} {\it Wind} ICME catalog (\href{https://wind.nasa.gov/fullcatalogue.php}{{\sf wind.nasa.gov/fullcatalogue.php}}) with circular flux rope model fitting.  Any CME arrivals at Earth (L1 spacecraft, e.g. $Wind$, $ACE$, and $DSCOVR$--Deep Space Climate Observatory), $STEREO$-A, and {\it STEREO-B} from the catalogues that had not been documented in DONKI were added to the database. In the case of any disagreement of arrival times between catalogues or DONKI, the in-situ data from $ACE$, $Wind$, $DSCOVR$, and {\it STEREO-A} and $B$ was analyzed. 

Complications arise in determining in-situ ICME arrivals for several reasons including (1) weak arrivals, (2) hybrid events including arrivals of stream interaction regions (SIRs) and CMEs, and (3) CME arrivals with uncertain sources.  The first complication occurs, as an example, when a CME has a slow speed and only creates a minimal jump in observed solar wind parameters.  In the second case of SIR/CME hybrids, it can be difficult to distinguish the CME arrival time from the jump caused by the SIR in the in-situ data. The third complication is likely to arise when there are multiple CMEs predicted to impact the same location at around the same time. The process of matching observed ICMEs to their source eruptions can be difficult.

Simulation information was retrieved from DONKI with a web-service application program interface (API). The DONKI database contains all CCMC space weather team WSA--ENLIL+Cone model runs, each linked to the CME measurement used as model input. Each CME is also linked to its ICME arrival observed at {\it STEREO-A}, {\it STEREO-B}, or Earth, verified with the ICME catalogues listed above. Arrivals at Mercury, Venus, and Mars are also recorded but are not used in this study. The predicted arrival times listed in DONKI are automatically obtained where the derivative of the simulated time series of the dynamic pressure crosses a threshold at each location. The forecaster may also add glancing blow CME arrival predictions by manually assessing the simulation contour plots for cases when a CME arrival is not automatically detected from the time series. The API returns a file in JavaScript Object Notation (JSON) format with all simulation and linked information. We developed automated python routines to process the JSON by CME, retrieving the ``CME analysis" measurements (velocity, direction, width) and the simulation results for each measurement. We calculated hit, miss, false alarm, and correct rejection statistics based on whether a predicted or observed arrival at Earth, {\it STEREO-A}, and/or {\it STEREO-B} is recorded (see Section \ref{sec:skillscores}). For the hits, CMEs both predicted and observed to arrive, we calculated the time difference, in hours, of the predicted and observed arrival.

This automated verification is complicated by the fact that multiple simulations are often performed for the same CME, either with updated CME input parameters or in conjunction with other CMEs. Since each CME analysis measurement can be flagged as either ``True" or ``False" in DONKI to indicate the best measurement, filtering out ``False" CME analyses and filtering by most final simulation results for each CME eliminates repeated CMEs from the error calculations. In cases when a CME with unchanging parameters is simulated multiple times (which occurs only when the CME is simulated with other CMEs), we only consider the first simulation of the CME that was performed. This helps eliminate any double counting of the same CME in different simulations that could skew our results.

Another problem can arise for simulations containing multiple CMEs impacting different locations. For example, if two CMEs are modeled together, ENLIL may output a predicted arrival at {\it STEREO-A} and at Earth. Only a visual examination of the simulation output can clearly show which CME is predicted to impact each location. The WSA--ENLIL+Cone model results stored in DONKI do not contain information that differentiates between which CME is predicted to impact each location, but this is a feature that CCMC plans to add. Since the predicted impacts of two CMEs modeled together are connected directly to the model and not also to their respective CMEs, this introduces some uncertainty to the skill score calculations because some correct rejections and misses may be counted multiple times for the same simulation, and some spurious false alarms will be counted. To test the impact of this database ambiguity, we excluded multiple simulations including the same CME (40\% of our sample) and most of the skill scores remained within the error bars reported (see discussion in Section \ref{sec:skillscores} and Figure \ref{fig:Hit_False_v3}). Note that this uncertainty does not impact our analysis of hits and CME arrival time error, as each CME is directly linked to an observed arrival in the database.

\section{CME Arrival Time Verification}
\label{sec:results}

\subsection{CME Arrival Time Prediction Errors}
\label{sec:error}

\begin{figure*}
    \includegraphics[width=\textwidth]{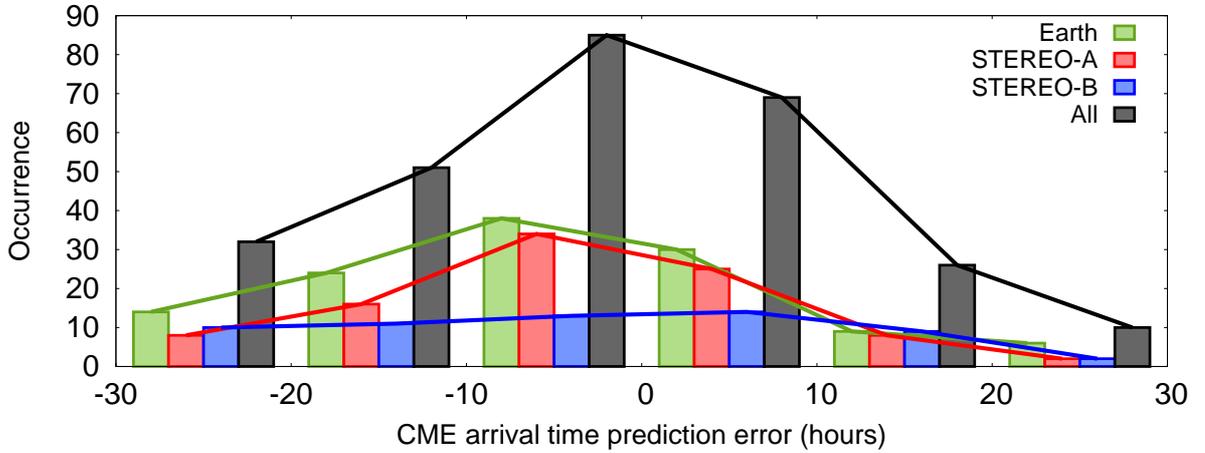}
    \caption{Distribution of CME arrival time prediction errors at Earth (green), {\it STEREO-A} (red), {\it STEREO-B} (blue), and all locations (black). The bins are as follow for each of the locations: [-30, -20], [-20, -10], [-10, 0], [0, 10], [10, 20], [20, 30]. The results for each location are distributed laterally within each bin space for clearer presentation.}
    \label{fig:error_distribution}
\end{figure*}

We computed the CME arrival time prediction error $\Delta t_{\rm err}=t_{\rm predicted}-t_{\rm observed}$ for hits at all three locations ({\it STEREO-A}, Earth, {\it STEREO-B}). A hit is defined when a CME is predicted and also observed to arrive. In the case of simulations that had arrival time prediction errors greater than 30 hours, the simulation was not counted as a hit, but as a miss. Figure \ref{fig:error_distribution} shows the histogram distribution of CME arrival time prediction errors in hours, with negative error indicating early prediction and positive error indicating late prediction. Overall at all locations, we found an average arrival time error of -4.0 hours showing a tendency for early prediction, and this can be seen in Figure \ref{fig:error_distribution}. The tendency for early predictions is -4.1 hours at Earth, -4.0 hours at {\it STEREO-A}, and -3.9 hours at {\it STEREO-B}. The distribution of errors at {\it STEREO-B} is flat compared to Earth and {\it STEREO-A}.

\begin{figure*}
    \includegraphics[width=\textwidth]{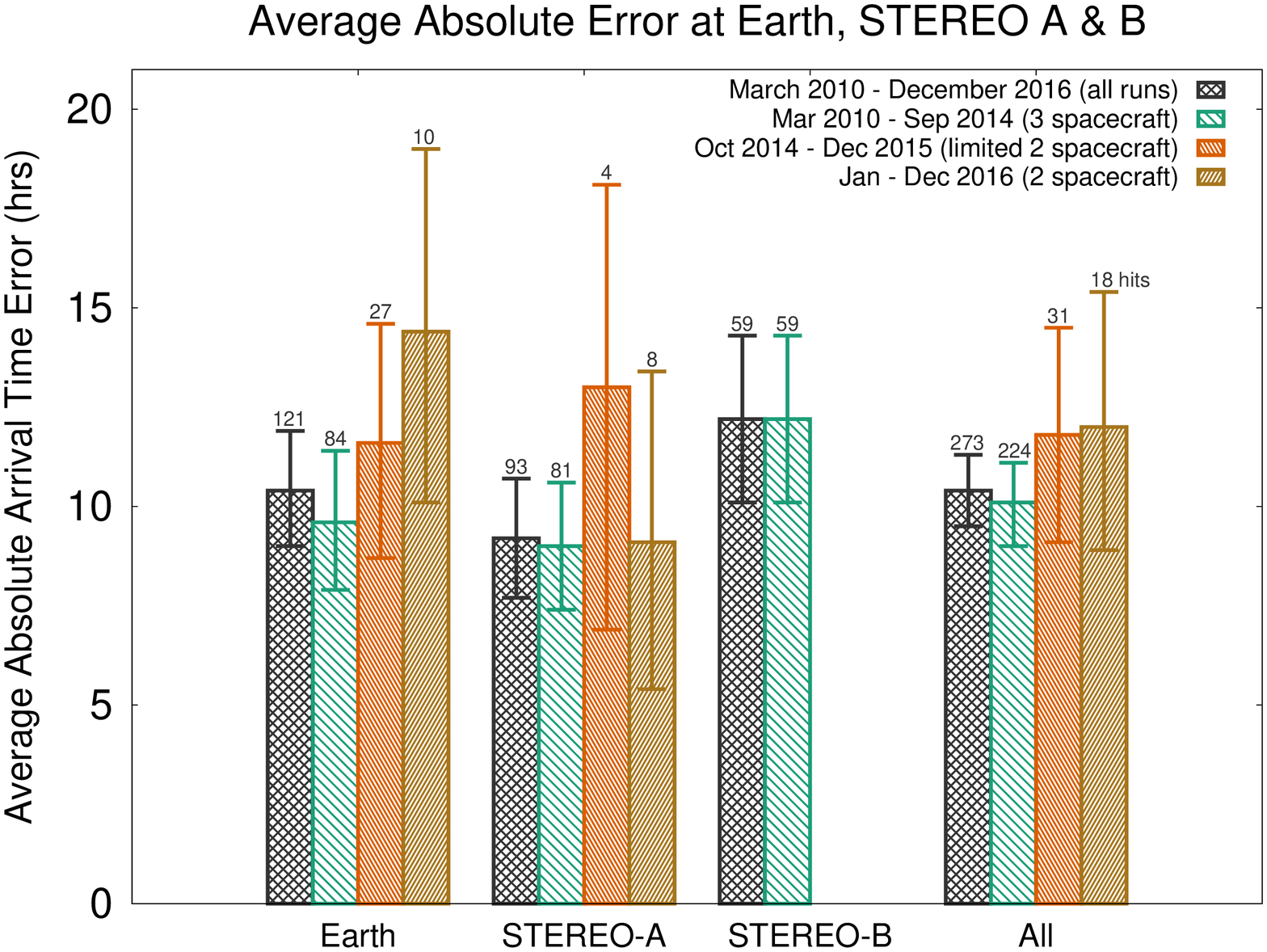}
    \caption{Average absolute error of CME arrival time predictions at Earth, {\it STEREO-A}, {\it STEREO-B}, and all together for four different time periods.}
    \label{fig:Absolute_errors_v3}
\end{figure*}


Figure \ref{fig:Absolute_errors_v3} shows the absolute arrival time error at each location and all together for the entire time period (black). We calculated 95\% confidence intervals using a bootstrapping method, resampling 10,000 times with replacement. We found an mean absolute error (MAE) of $10.4^{+1.5}_{-1.4}$ hours at Earth, $9.2^{+1.5}_{-1.4}$ hours at {\it STEREO-A}, $12.2\pm2.1$ hours at {\it STEREO-B}, and $10.4\pm0.9$ hours at all locations considered together.  The slightly increased arrival time error at {\it STEREO-B} may be due to the model being initialized by the oldest magnetogram information among the three locations (however the error bars at each location overlap).   As  discussed earlier, prediction errors were computed for hits, defined when the CME arrival time error is less than 30 hours.  To examine the effect of the threshold used to define the hit, we varied the threshold by decreasing it to 24 and 18 hours. As expected, the MAE at all locations decreases to $8.9\pm0.8$ and $7.4\pm0.7$ hours respectively.

\begin{figure*}
    \includegraphics[width=\textwidth]{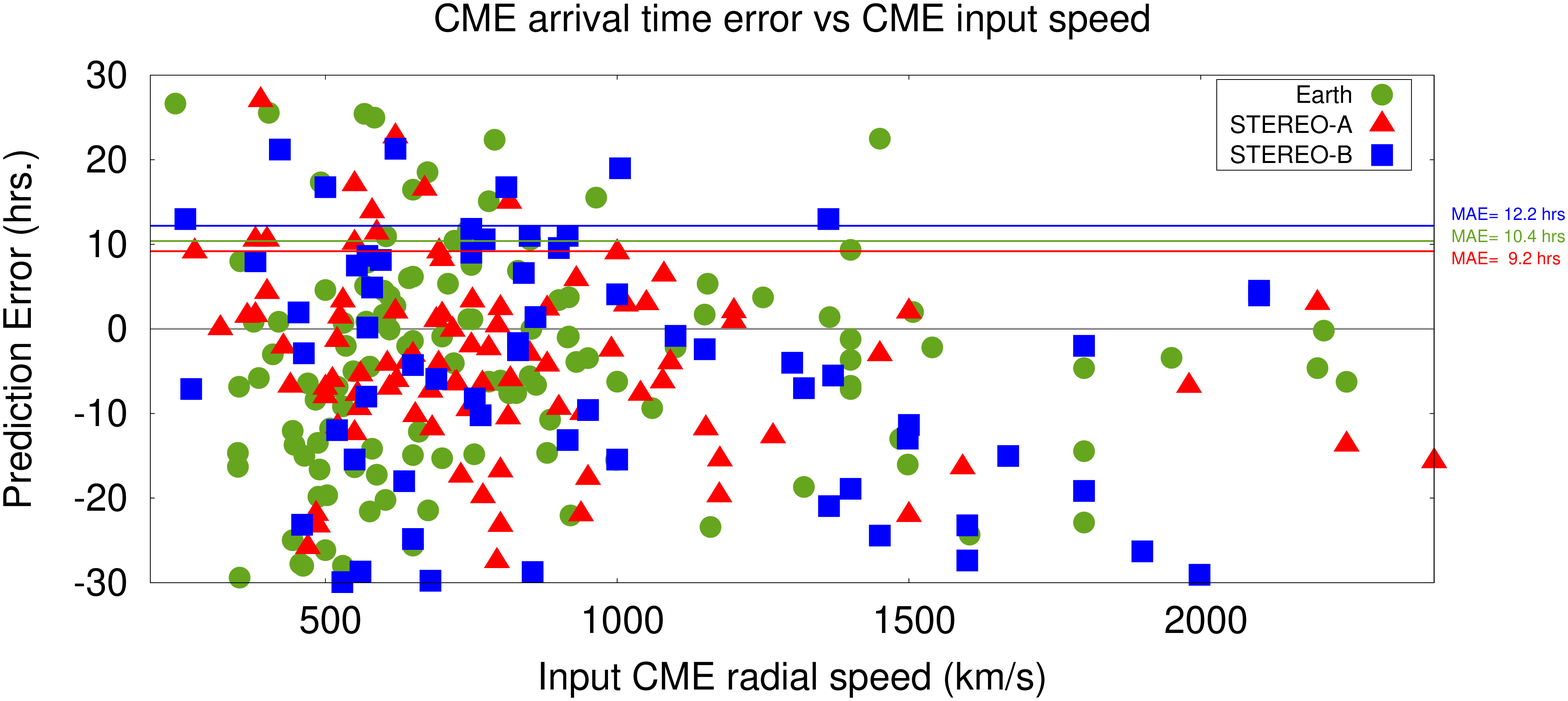}
    \includegraphics[width=\textwidth]{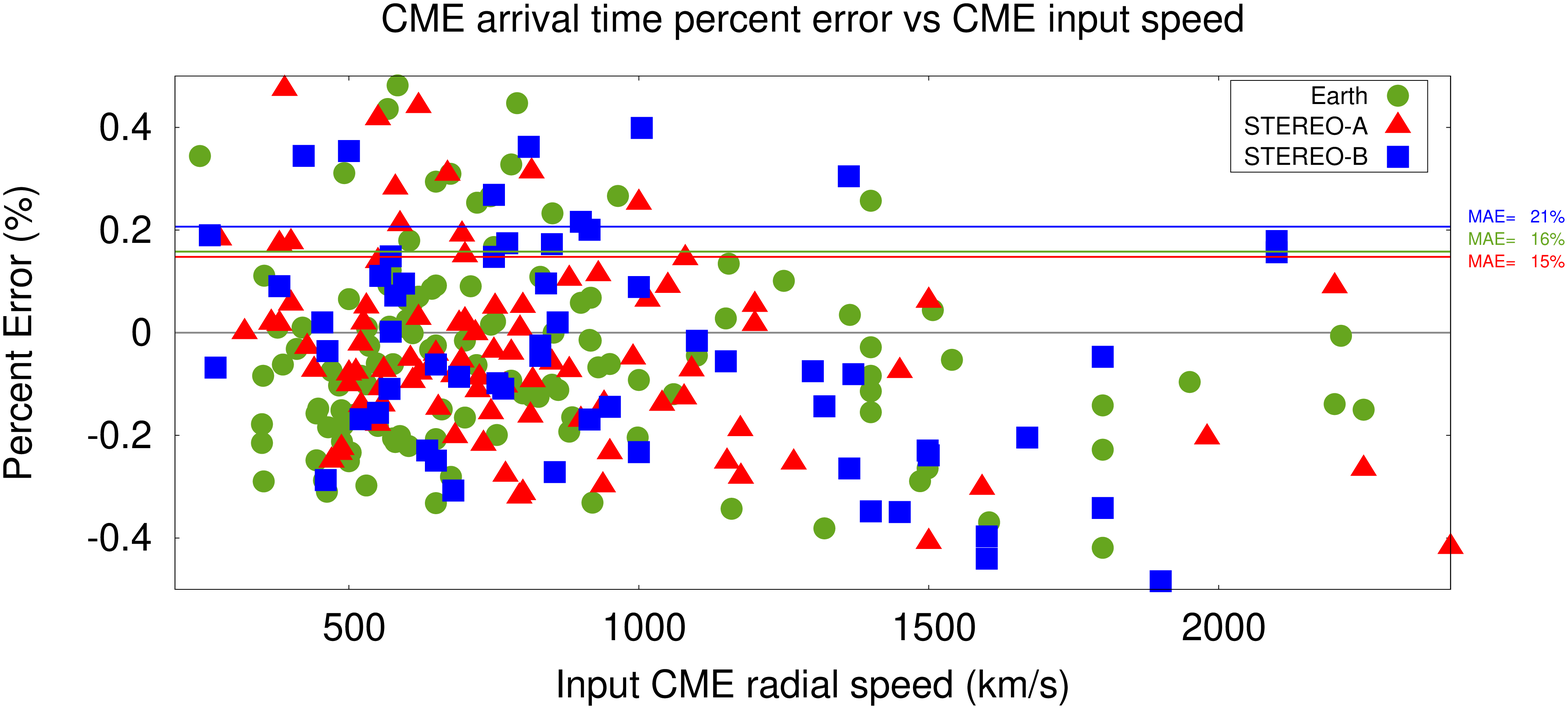}
    
    \caption{CME arrival time prediction error versus CME input radial speed (top) and CME arrival time percent error versus CME input radial speed (bottom).}
    \label{fig:error_velocity}
\end{figure*}

\begin{figure*}
    \includegraphics[width=\textwidth]{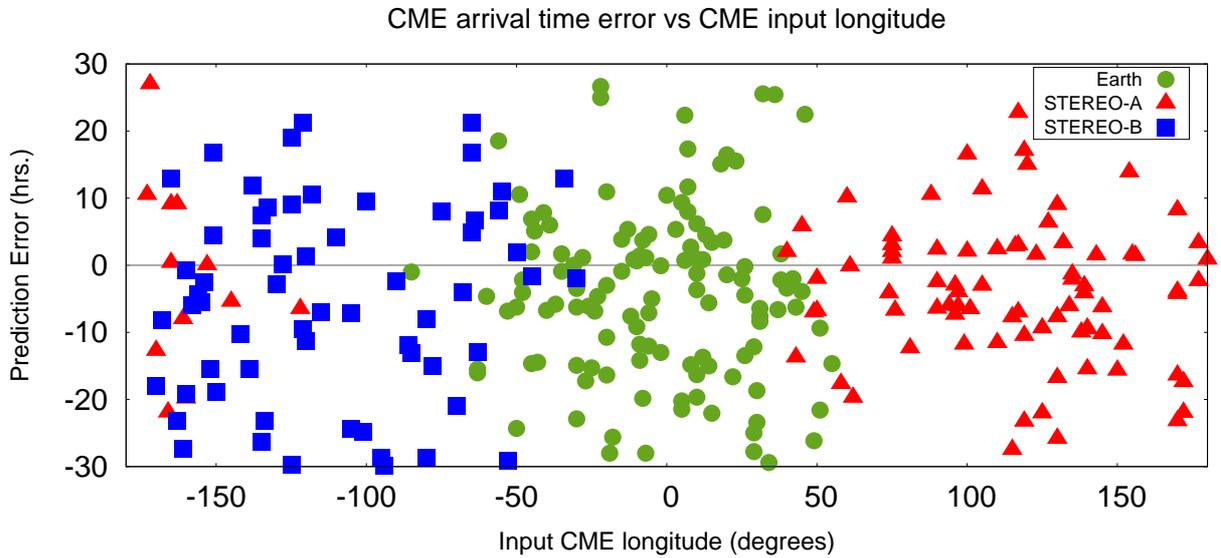}
    \caption{CME arrival time prediction errors versus CME longitude.}
    \label{fig:error_lon}    
\end{figure*}

\begin{figure}
\includegraphics[width=0.5\textwidth]{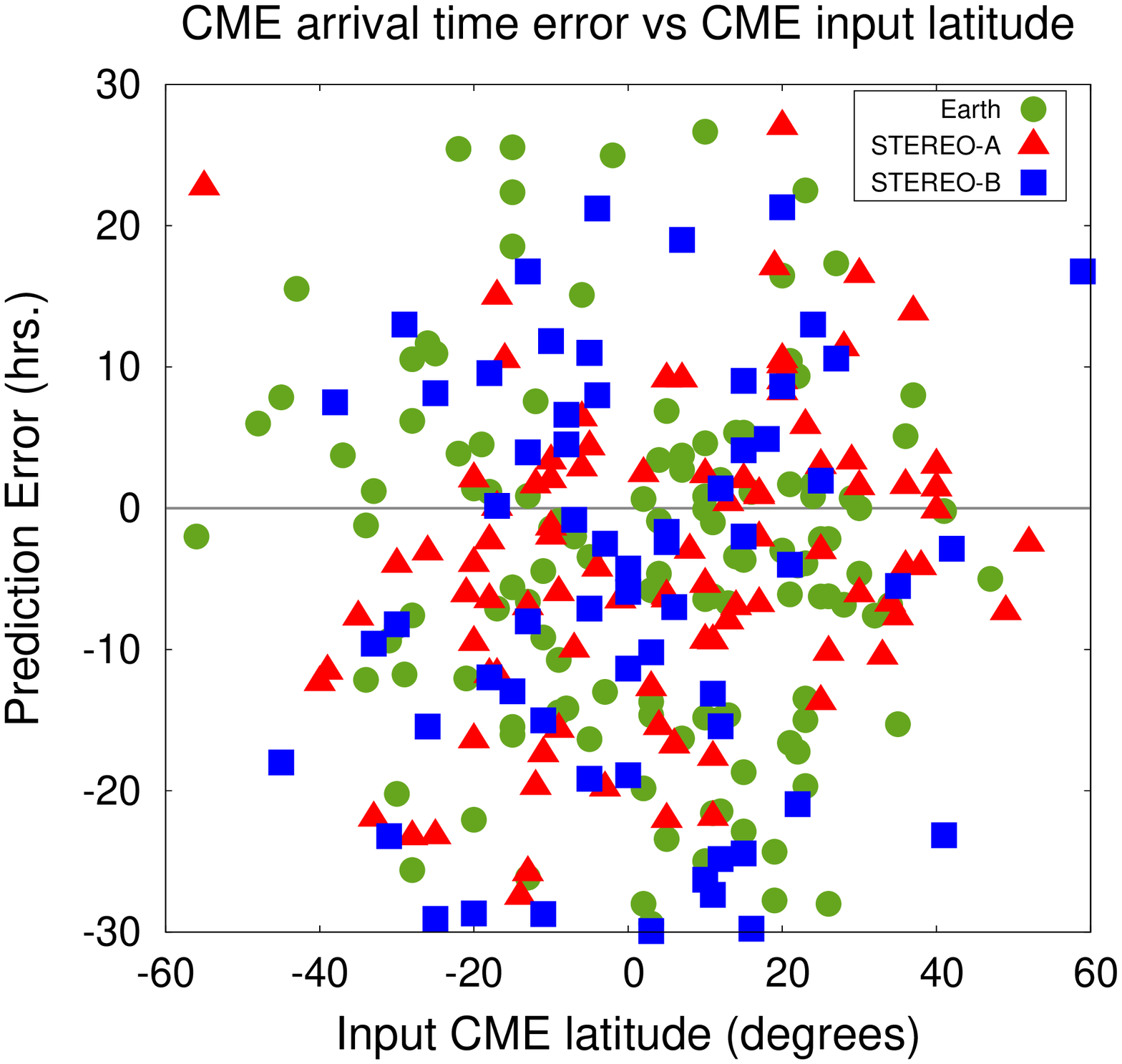}
    \includegraphics[width=0.5\textwidth]{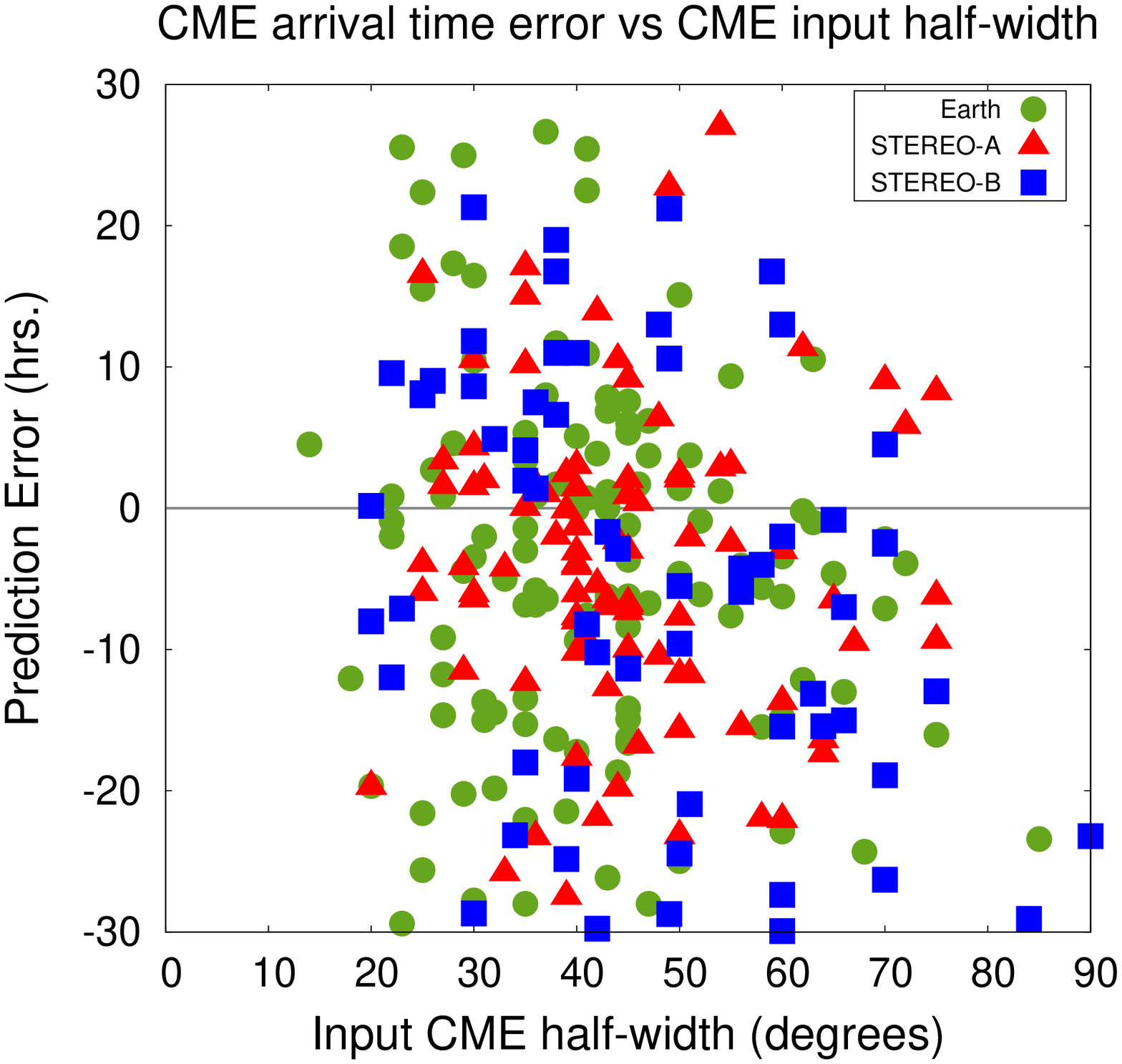}
        \caption{CME arrival time prediction errors versus input CME latitude (left) and CME half-width (right).}
    \label{fig:error_lat_width}
\end{figure}

In Figure \ref{fig:error_velocity} (top) we plot the prediction error in hours against the input CME radial speed in km\,s$^{-1}$. Looking at CMEs with input speeds below 1000 km\,s$^{-1}$, errors are scattered. However, with CMEs with speeds over 1000 km\,s$^{-1}$, most arrival time predictions are early, similar to the results found by \cite{mays2015}. This could be a sign of the modeled CME having too much momentum as defined by a combination of the input speed and half-width (which is related to the modeled CME mass). The overestimation of the modeled CME velocity compared to in-situ observed values is also due to the modeled CME having a lower magnetic pressure than is observed in typical magnetic clouds. If we exclude CME input speeds below approximately 700 km\,s$^{-1}$ the MAE becomes $8.5$ hours at Earth, $8.4$ hours at {\it STEREO-A}, $12.2$ hours at {\it STEREO-B}.  Figure \ref{fig:error_velocity} (bottom) shows the CME arrival time percent error versus the input CME radial velocity. The percent error is calculated as the ratio of the prediction error to the transit time from 21.5 R$_{\odot}$ to the detecting spacecraft, and this figure shows similar trends. In investigating any correlation between arrival time error and CME input parameters, no significant correlation was found with latitude, longitude, or width.  The distribution of input longitudes in Figure \ref{fig:error_lon} roughly correlates with the detecting spacecraft's location. CME input latitudes in Figure \ref{fig:error_lat_width} (left) are scattered between +60$^{\circ}$ and -60$^{\circ}$, as expected for CMEs that are detected at spacecraft within the ecliptic plane.  There is no obvious correlation between error and CME input width in Figure \ref{fig:error_lat_width} (right).

\subsection{Different Spacecraft Viewpoint Time Periods}
\label{sec:subsets}
Figure \ref{fig:Absolute_errors_v3} also shows the error at each location for three different time periods, useful for comparing the arrival time errors during periods of two or three observing spacecraft. The full time period of March 2010--December 2016 is shown in black and was discussed in the previous section. The March 2010--September 2014 period had three coronagraph viewpoints observing CMEs, shown in green. {\it STEREO-A} longitudes ranged from 65$^{\circ}$ to 168$^{\circ}$ and {\it B} -71$^{\circ}$ to -161$^{\circ}$ during this time period. After September 2014, communication with {\it STEREO-B} was lost, leaving only two coronagraph viewpoints. From October 2014 through December 2015 (orange), just after the {\it STEREO-B} communication loss, {\it STEREO-A} underwent sidelobe operations, resulting in very limited {\it STEREO-A} coverage during this period and effectively only a single spacecraft was available for CME measurement. Normal {\it STEREO-A} operations resumed in the last time period of January--December 2016 (brown), there are only 18 hits during this period. {\it STEREO-A} longitudes ranged from -165$^{\circ}$ to -143$^{\circ}$ during January--December 2016, which was far from Earth and reduced the effectiveness of dual coronagraph observations.

The average absolute arrival time error at all locations from March 2010--September 2014 with 224 hits  (green: three spacecraft) was $10.1\pm1.0$ hours. For the 31 hits during the period of {\it STEREO-A} sidelobe operations and {\it STEREO-B} communication loss (orange: single-viewpoint in effect), the mean absolute arrival-time prediction error is $11.8^{+2.7}_{-2.6}$ hours at all locations. In comparing these two time periods, there is a reduction in skill of $1.7^{+2.9}_{-2.8}$ hours. However, due to the small sample size of only 31 hits in the October 2014--December 2015 (orange) {\it STEREO-A} sidelobe operation period, the 95\% error bars overlap.  For the 18 hits after the end of {\it STEREO-A} sidelobe operations, but still without {\it STEREO-B} communication from January--December 2016 (brown: two spacecraft), the mean absolute arrival-time error is $12.0^{+3.4}_{-3.2}$ hours. If we combine the two spacecraft viewpoint periods from October 2014--December 2016 (orange and brown) the error is $11.9^{+2.1}_{-2.0}$ hours, with more events (49) reducing the error bar.  Compared to the three spacecraft period (green) the difference becomes $1.8^{+2.3}_{-2.2}$ hours, still overlapping. The overlapping error bars for these time periods due to the small number of hits in the two-viewpoint time period shows that this result does not have 95\% statistical significance. However, if we reduce our bootstrapped confidence interval to 60\%, the difference in arrival time errors for three-viewpoint (green) vs single-viewpoint in effect (orange) time period is $1.7\pm1.2$ hours (without overlap). This shows a trend for multi-view coronagraph observations improving CME arrival time forecast accuracy with 60\% confidence and that more events are needed to show this with 95\% statistical significance. 

Similarly, considering only Earth hits, the difference in arrival time errors for the three-viewpoint (green; MAE=$9.6\pm0.8$ hrs) vs single-viewpoint in effect (orange; MAE=$11.6\pm1.4$ hrs) time period is $2.0\pm0.6$ hours if the confidence interval is reduced to 60\% (the 95\% confidence interval overlaps). For the 10 hits at Earth from January--December 2016 the MAE=$14.4^{+4.5}_{-4.3}$ hours. This increased error compared to the sidelobe period could indicate that the {\it STEREO-A} viewpoint on the farside did not add to greater accuracy in CME measurement. If we combine the two-spacecraft viewpoint periods of October 2014--December 2016 (orange and brown) the error difference with the three spacecraft period is $2.7\pm1.9$ hours (75\% confidence interval). When considering only STEREO-A arrivals, the difference in arrival time error between the time period with three viewpoints (green) and the time period with an effective single viewpoint (orange) is $4.1^{+4.87}_{-6.7}$ hours (with overlap).  This large error may suggest that there is a reduction in CME arrival time accuracy when there is only a farside coronagraph (Earth viewpoint) and an extreme ultraviolet instrument (STEREO-A viewpoint). However, there are only 4 hits in this reduced time period, so this is not conclusive.
 
For completeness we have examined all of the CME simulations included in our verification study.  This included glancing blow CMEs that are generally more difficult to predict.  In the DONKI database predicted glancing blows are manually entered into the system by the forecaster and are flagged as such.  We performed the same analysis as Section \ref{sec:error} excluding predicted glancing blow arrivals, which reduced the hits at all locations from 273 to 183 and found an average absolute CME arrival time error at all locations of $10.2\pm1.1$ hours and $9.5^{+1.7}_{-1.6}$ hours at Earth. As expected, the error decreases when glancing blows are excluded, but we only find a slight decrease.

\subsection{CME Arrival Time Skill Scores}
\label{sec:skillscores}

For the purpose of evaluating our forecasting performance, we defined each simulation as a hit, miss, false alarm, or correct rejection.  A hit is an event forecast to occur that did occur. A miss is an event forecast not to occur, but did occur. A false alarm is an event forecast to occur that did not occur. A correct rejection is an event forecast not to occur that did not occur. This is summarized in Table \ref{table:contingencydef}. In the case of simulations that had a difference in prediction time and observed arrival time greater than 30 hours, the simulation was not counted as a hit, but as a miss. A variety of skill scores were calculated to evaluate model performance as defined in Table \ref{table:skillscoresdef}.  The success and false alarm ratios are conditioned on the forecasts (given that an event was forecast, what was the observed outcome?) and represent the fraction of predicted events that were observed, and were not observed, respectively. The accuracy score is the overall fraction of correct forecasts. The bias score indicates if the model is overforecasting or underforecasting by measuring the ratio of the frequency of forecast events to the frequency of observed events, and ranges from -1 to 1. The probability of detection (POD, or hit rate) and probability of false detection (POFD, or false alarm rate) are conditioned on the observations (given that an event was observed, what is the corresponding forecast?) as the fraction of observed events that were predicted and fraction of incorrect observed non-events, respectively. The Hanssen and Kuipers discriminant ($HK = POD - POFD$), or true skill statistic, or Pierce's skill score, ranges from -1 to 1, with 1 being a perfect score and 0 is no skill. The HK discriminant measures the ability of the forecast to discriminate between two alternative outcomes and does not rely on climatological event frequency \citep{jolliffe2011}.

\begin{table}
\caption{Contingency Table.}             
\label{table:contingencydef}      
\centering                          
\begin{tabular}{c c c}        
\hline\hline                 
 & Observed Arrival & No Observed Arrival \\    
\hline                        
 Predicted Arrival & Hit (H) & False Alarm (FA) \\      
 No Predicted Arrival & Miss (M) & Correct Rejection (CR)\\ 
\hline                                   
\end{tabular}
\end{table}

\begin{table}
\caption{Brief description of skill scores derived from the contingency table. The false alarm rate is also known as the probability of false detection (POFD) and the hit rate as the probability of detection (POD).}
\label{table:skillscoresdef}
\centering
\begin{tabular}{l c c l}
\hline\hline 
Skill Score & Equation& Perfect Score & Definition  \\ 
\hline
   Success Ratio        & $H/(H + FA)$     & 1 & Fraction of correct predicted arrivals. \\
   False Alarm Ratio        & $FA/(H + FA)$    & 0 & Fraction of incorrect predicted arrivals.  \\
  
   Accuracy             & $(H+CR)/Total$   & 1 & Fraction of correct forecasts.\\
   Bias Score           & $(H + FA)/(H + M)$ & 1 & Ratio of predicted arrivals to observed arrivals, \\
                        &                           &   & $<1=$ underforecast; $>1=$ overforecast \\
   Hit Rate (POD)       & $H/(H + M)$      & 1 & Fraction of observed arrivals that were predicted.  \\
    False Alarm Rate (POFD)       & $FA/(CR + FA)$    & 0 & Fraction of incorrect observed non-arrivals  \\
    Hanssen \& Kuipers& {\it \footnotesize HK=POD-POFD}& 1 & Forecast ability to discriminate between\\
    discriminant & &  & observed event occurrence from non-occurrence\\
\hline
\end{tabular}
\end{table}

Table \ref{table:contingencydata} shows the number of hits, misses, false alarms, and correct rejections for the real-time simulations run at the CCMC from March 2010 to December 2016. The greater than symbol before the misses at {\it STEREO-A} and {\it STEREO-B} indicates that there are a greater number of misses than we could fully confirm at these locations. This is due to the ICME catalogues for {\it STEREO-A} and {\it STEREO-B} having not been updated past 2014 at the time of our analysis. They have since been updated through 2016 at the time of this writing.

\begin{table}
\caption{Hit, miss, false alarm, and correct rejection rates for the WSA--ENLIL+Cone model for the period March 2010--December 2016.}             
\label{table:contingencydata}      
\centering                          
\begin{tabular}{c c c c c}        
\hline\hline                 
 & Earth & {\it STEREO-A} & {\it STEREO-B} & All \\    
\hline                        
Hits & 121 & 93	& 59 & 273 \\
False Alarms & 180 & 127 & 95 & 402 \\
Misses & 106 & $>$110 & $>$75 & $>$291 \\
Correct Rejections & 1293 & 1393 & 1017 & 3703 \\
\hline                                   
\end{tabular}
\end{table}

\begin{figure*}
    \includegraphics[width=\textwidth]{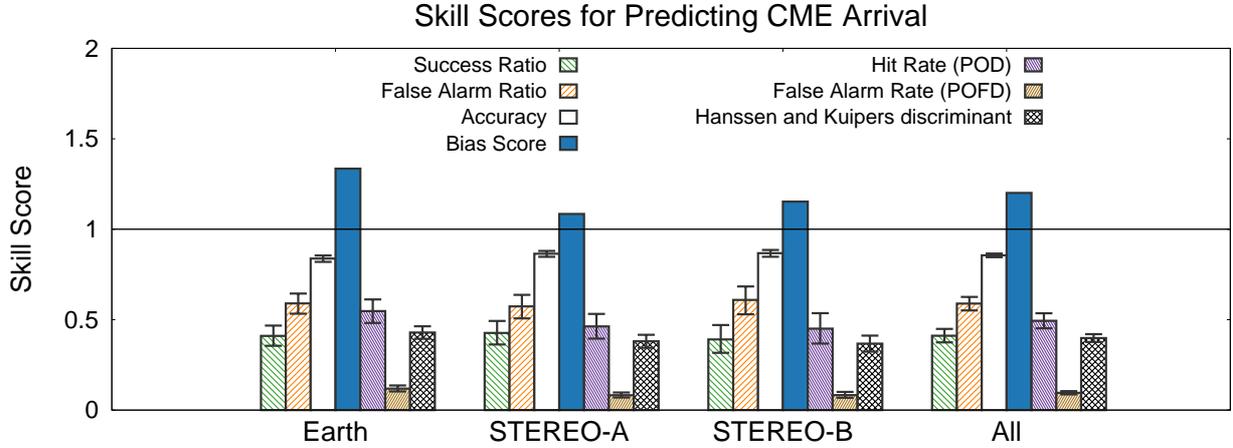}
    \caption{Success ratio, false alarm ratio, accuracy score, bias score, POD, POFD, and HK (defined in Table \ref{table:skillscoresdef}) of total modeled CME events which predict hits at Earth, {\it STEREO-A}, {\it STEREO-B}, and all locations combined. Error bars derived from \cite{wilks2011}.}
    \label{fig:Hit_False_v3}
\end{figure*}

Figure \ref{fig:Hit_False_v3} shows the calculated skill scores discussed in Table \ref{table:skillscoresdef}.  With the exception of the bias score, the skill scores do not vary greatly between locations. At all locations over the whole time period, the hit rate, or probability of detection (POD) is 0.50, indicating that half of the observed arrivals were correctly predicted. The false alarm rate, or probability of false detection (POFD), is 0.10 and is the fraction of observed non-arrivals that were incorrectly predicted.  The Hanssen and Kuipers discriminant ($HK = POD - POFD$) of 0.39 represents how well the prediction is able to separates arrivals from non-arrivals. On the other hand, the success ratio is 0.40 and the false alarm ratio is 0.60. The success and false alarm ratios represent the fraction of predicted CME arrivals that were observed, and were not observed, respectively. The success ratio is less than the false alarm ratio, and both scores are far from the perfect scores of 1 and 0, respectively, showing poor skill for these scores which are conditioned on predictions.  However, due to spurious false alarms possible from the DONKI database ambiguity discussed in Section \ref{sec:method}, when we excluded all simulations of more than one CME (40\% of our sample) the success and false alarm ratios become 0.55 and 0.45 respectively, showing some minor improvement. The database ambiguity did not effect any other skill scores to within the error bars shown in Figure \ref{fig:Hit_False_v3}. The accuracy score at all locations is 0.85 and is closer to the perfect score than the success ratio or hit rate because it is the fraction of correct forecasts overall and the correct rejections are the majority of our forecasts.

The bias score is the ratio of the frequency of forecast events to the frequency of observed events and shows whether the model is underforecasting or over-forecasting. We find a bias score of 1.2 for all locations considered together, revealing a tendency to overforecast CME arrivals.  There is a slight bias of 1.1 at {\it STEREO-A} and {\it STEREO-B}, but at Earth there is the highest bias for overforecasting with a bias score of 1.3.  This may be due to CCMC space weather team assessing glancing blows in simulation results that were not automatically detected as CME arrivals more often for Earth-directed events, as they have the potential for geoeffectiveness. There is a human bias for over-predicting CME arrivals at Earth by modeling more events. \cite{millward2014} notes that forecaster judgment plays a significant role in event selection and that operational model performance assessments will include the more-difficult-to-forecast glancing blow events compared to non-real-time assessments.

\section{$K_{\rm P}$ Prediction Verification}
\label{sec:kp}
\begin{figure*}
    \includegraphics[width=\textwidth]{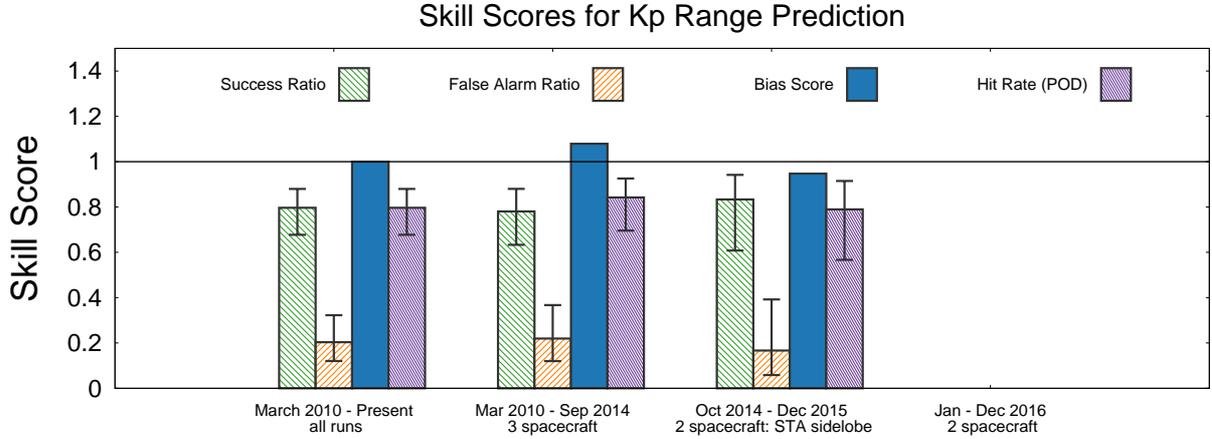}
    \caption{Success ratio, false alarm ratio, bias score, and hit rate skill scores based on whether the observed $K_{\rm p\,{max}}$ falls within the predicted $K_{\rm P}$ range, grouped by forecast time period.}
    \label{fig:Hit_False_Skill_Scores_Kp_v1}
\end{figure*}

The geomagnetic three-hour planetary $K$ index ($K_{\rm P}$) is derived from data from ground-based magnetometers, categorizes geomagnetic activity on a scale from 0 to 9, with 0 being the lowest amount of geomagnetic activity and 9 being the highest \citep{bartels1939,rostoker1972,menvielle1991}. The time series that the WSA--ENLIL+Cone model outputs at Earth is used as input to a formula derived from the \cite{newell2007} coupling function to provide a $K_{\rm P}$ forecast time series for three magnetic field clock-angle scenarios of 90$^{\circ}$ (westward), 135$^{\circ}$ (southwestward), and 180$^{\circ}$ (southward).  Because ENLIL-modeled CMEs do not contain an internal magnetic field and the magnetic field amplification is caused mostly by plasma compression, only the magnetic- field magnitude is used and the three magnetic field clock-angle scenarios are assumed. This provides a simple estimate of three possible maximum values of each time series that the $K_{\rm P}$ index might reach following arrival of the predicted CME shock/sheath. For the forecast, the $K_{\rm P}$ estimates are rounded to the nearest whole number \citep{mays2015}. With all three clock-angle forecasts, we determined a range from the minimum to the maximum predicted $K_{\rm P}$ and compared to the maximum observed $K_{\rm P}$ for the three days following the CME arrival. If the maximum observed $K_{\rm P}$ within three days after the observed CME arrival fell within the predicted $K_{\rm P}$ range, we counted a hit. If the max observed $K_{\rm P}$ was above the forecast range, we counted a miss. If the max observed $K_{\rm P}$ was below the forecast range, we counted a false alarm. For example, if the predicted $K_{\rm p\,{max}}$ for the three clock angle scenarios is 5, 7, 7, and the observed $K_{\rm P}$ is 6, this is counted as a hit because 6 falls within the predicted $K_{\rm p\,{max}}$ range of 5--7. The resulting contingency table for $K_{\rm P}$ range is: 47 hits, 12 false alarms, and 12 misses. There are no correct rejections in this analysis, so we could not calculate an accuracy score. Thus, Figure \ref{fig:Hit_False_Skill_Scores_Kp_v1} shows all skill scores from Table \ref{table:skillscoresdef} except for  accuracy, for the different time periods defined in Section \ref{sec:error}. There were only two $K_{\rm P}$ predictions for the January to December 2016 time period, so skill scores are not shown. For the entire time period, we found a success ratio of 0.80 and a false alarm ratio of 0.20, meaning that a high fraction of predicted $K_{\rm P}$ ranges contained observed $K_{\rm P}$ value and a low fraction of predicted $K_{\rm P}$ ranges did not contain the observed value. We found a bias score of 1.0, indicating that we are neither under- or over-forecasting the $K_{\rm P}$ range. The hit rate (POD) of 0.80 shows a high fraction of the observed $K_{\rm P}$ values were within the $K_{\rm P}$ range.

\begin{figure*}
    \includegraphics[width=\textwidth]{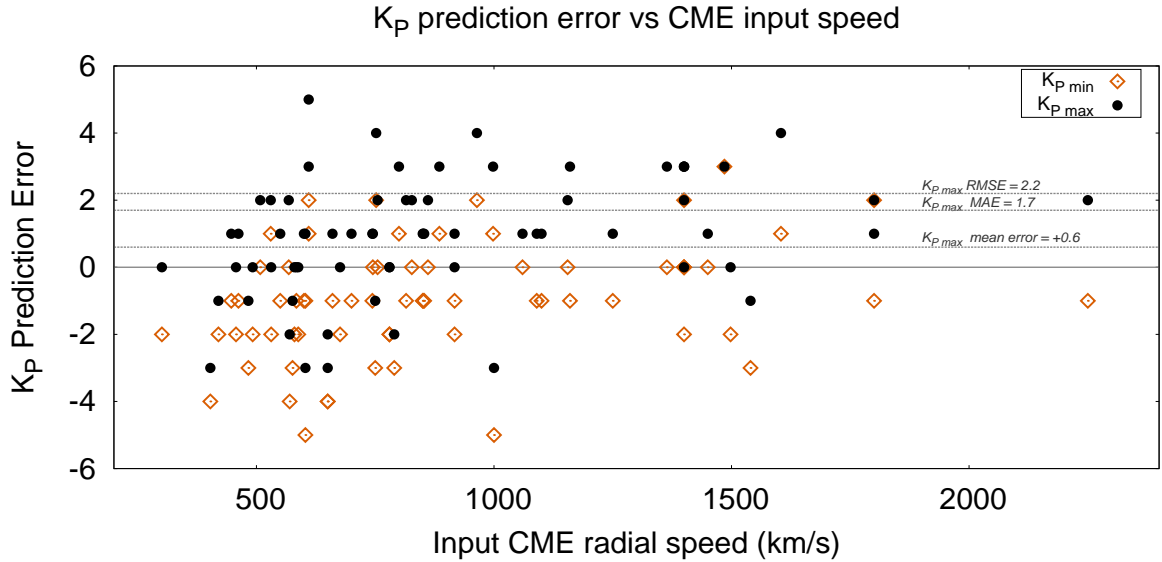}
    \caption{$K_{\rm p\,{max}}$ and $K_{\rm p\,{min}}$ prediction error plotted against the CME input speed, at times showing large errors given the range of the index from 0 to 9. There is a general overprediction of $K_{\rm p\,{max}}$ for CME input speeds above $\approx$\,1000 km\,s$^{-1}$, whereas the $K_{\rm p\,{min}}$ forecast tends underpredict below $\approx$\,1000 km\,s$^{-1}$.}
    \label{fig:kpspeed}
\end{figure*}

\begin{figure*}
    \includegraphics[width=\textwidth]{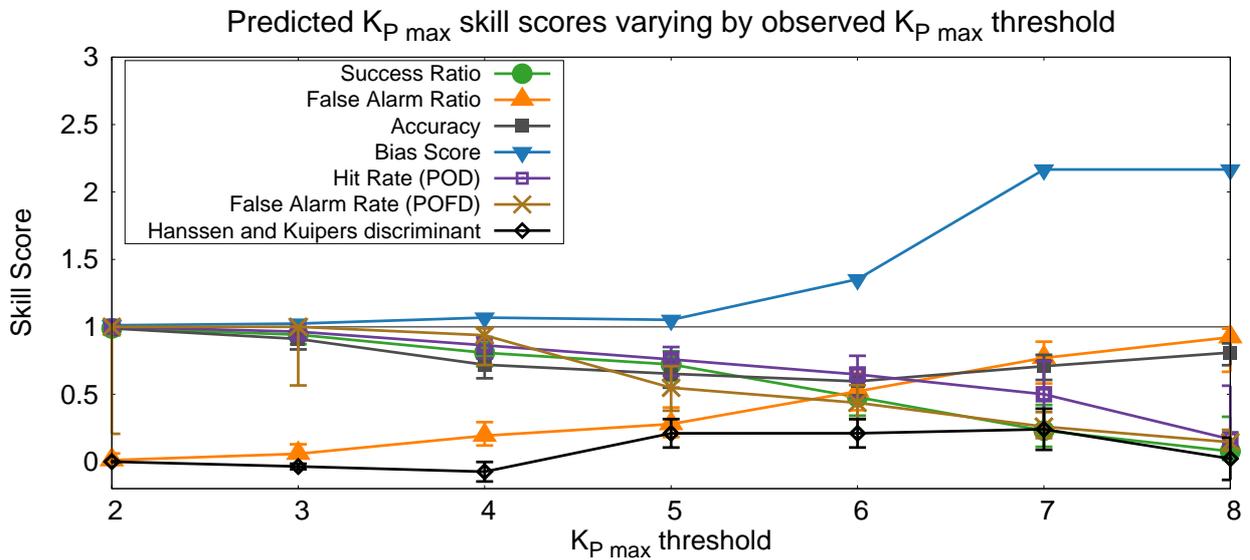}
    \caption{Success ratio, false alarm ratio, accuracy, bias score, and hit rate skill scores for $K_{\rm P\,{max}}$ forecasts as a function of different thresholds.}
    \label{fig:ss_threshold}
\end{figure*}

The $K_{\rm P}$ prediction error can be calculated by first taking the maximum of the three $K_{\rm p\,{max}}$ predictions as a single prediction (this is generally 180$^{\circ}$). For example, for the earlier prediction of $K_{\rm p\,{max}}$=5,7,7, the single maximum $K_{\rm p\,{max}}$ prediction is 7. For all $K_{\rm P}$ predictions, the $K_{\rm p\,{max}}$ mean error is 0.6$\pm0.4$, the mean average error is 1.7$\pm0.3$, and the RMSE is 2.2$\pm0.3$. The error bars are computed using a similar bootstrapping method as used for the arrival time error. In Figure \ref{fig:kpspeed}, $K_{\rm p\,{max}}$ and $K_{\rm p\,{min}}$ prediction errors are plotted against the CME input speed, at times showing large errors given the range of the index from 0 to 9. We find that there is a general overprediction of $K_{\rm p\,{max}}$ for CME input speeds above $\approx$\,1000 km\,s$^{-1}$, similar to the results found by \cite{mays2015}. This overprediction could be due to an overestimation of the CME dynamic pressure at Earth for faster CMEs, as the inserted CMEs have a lower magnetic pressure than is observed and from the approximation of the CME as a cloud with homogeneous density. We also see that the $K_{\rm p\,{min}}$ forecast underpredicts $K_{\rm P}$ for CME speeds under $\approx$\,1000 km\,s$^{-1}$, likely when the observed $K_{\rm P}$ is caused by magnetic cloud passage, and not the shock/sheath.

Instead of using the forecast $K_{\rm P}$ range, a threshold was used to compute the $K_{\rm p\,{max}}$ prediction skill scores in Figure \ref{fig:ss_threshold}. For example, for a threshold of $K_{\rm P}$ = 6, a hit occurs when the forecast $K_{\rm p\,{max}}$ and the observed $K_{\rm p\,{max}}$ are at or above 6. A correct rejection is counted when the forecast $K_{\rm p\,{max}}$ and the observed $K_{\rm p\,{max}}$ are both below 6. A false alarm is counted when the forecast $K_{\rm p\,{max}}$ is above or at 6 and the observed $K_{\rm p\,{max}}$ is below 6. A miss is counted when the forecast $K_{\rm p\,{max}}$ is below 6 and the observed $K_{\rm p\,{max}}$ is at or above 6. We varied the threshold between 2 and 8 in Figure \ref{fig:ss_threshold} and computed the resulting skill scores.  The figure shows the success ratio, false alarm ratio, accuracy, bias score, hit rate, false alarm rate, and HK as a function of threshold. The skill scores generally decrease in performance as the $K_{\rm p\,{max}}$ threshold increases, except the false alarm ratio, and HK. With events categorized as hits at low thresholds shifting to correct rejections at higher thresholds, the total number of hits is decreasing. The success ratio, accuracy score, bias score, hit rate, and false alarm rate are all heavily dependent on the number of hits. Only the accuracy score remains relatively unchanged, since it depends on the total number of events (unchanging). This figure shows that the $K_{\rm p\,{max}}$ prediction performs best (where most skill scores perform the best) for an observed $K_{\rm p\,{max}}$ of 5 and above, after which the $K_{\rm p\,{max}}$ is overforecast (bias score) and the false alarm ratio increases.

\section{Summary and Discussion}
\label{sec:conclusions}
In this article we evaluate the performance of the WSA--ENLIL+Cone model installed at the CCMC and executed in real-time by the CCMC space weather team from March 2010 to December 2016. The simulations included over 1,800 CMEs and 273 of these were categorized as hits---CMEs that were both observed and predicted to arrive. We computed the CME arrival time prediction error $\Delta t_{\rm err}=t_{\rm predicted}-t_{\rm observed}$ for hits at three locations: Earth, {\it STEREO-A} and {B}. The average absolute error of the CME arrival time was found to be $10.4^{+1.5}_{-1.4}$ hours at Earth, $9.2^{+1.5}_{-1.4}$ hours {\it STEREO-A}, $12.2\pm2.1$ hours at {\it STEREO-B}, and $10.4\pm0.9$ hours at all locations considered together. These errors are comparable to arrival time error results from other studies using WSA-ENLIL or other models (see Section \ref{sec:intro}).  At all locations over the whole time period, the hit rate is 0.50, indicating that half of the observed arrivals were correctly predicted. The false alarm rate is 0.10 and represents the fraction of observed non-arrivals that were incorrectly predicted.

Overall, at all locations, we found an average arrival time error of -4.0 hours showing a tendency for early predictions.  Sources of CME arrival time error include input CME parameters, ambient solar wind prediction accuracy, previous CMEs, input magnetogram limitations/uncertainties, model ambient parameters, and model CME parameters. For example, \cite{temmer2017} found that interplanetary space takes about 2--5 days to recover to normal background solar wind speed conditions. We are exploring methods to quantify the effect of the ambient solar wind prediction accuracy on CME arrival time, but this is beyond the scope of this article. The tendency for early predictions can also arise from CME input speeds, as measured from coronagraph data, that are too high from measurement of the shock instead of the main CME driver \citep{mays2015b}.  A parametric study by \cite{mays2015} found that larger CME half widths also have a small tendency to produce early arrivals.  They also report that reducing the ENLIL model parameter of CME cloud density ratio [{\sf{dcld}}] may help to predict later arrivals. Overall, the parametric case study shows that after the CME input speed, the cavity ratio [{\sf{radcav}}] (radial CME cavity width/CME width) and density ratio assumed in ENLIL have the greatest effects on the predicted CME arrival time. 

For Earth-directed CMEs using the WSA--ENLIL+Cone model outputs to compute an estimate of the geomagnetic $K_{\rm P}$ index we found a mean $K_{\rm p\,{max}}$ error of 0.6$\pm0.4$, a mean average error of 1.7$\pm0.3$, and RMSE of 2.2$\pm0.3$, and the $K_{\rm p\,{max}}$ is generally overpredicted for CME input speeds above $\approx$\,1000 km\,s$^{-1}$. The $K_{\rm p\,{max}}$ overprediction could be due to an overestimation of the CME dynamic pressure at Earth for faster CMEs, as the inserted CMEs have no magnetic pressure other than from the ambient field.

Verification of the single-spacecraft (in effect) period October 2014--December 2015 (without {\it STEREO-B} and with reduced {\it STEREO-A} coverage) shows an increase in CME arrival time error of $1.7^{+2.9}_{-2.8}$ hours.  Because of overlapping error bars for these time periods due to the small number of hits in the single-viewpoint (in effect) time period this result does not have 95\% statistical significance, but instead 60\% ($1.7\pm1.2$). Nevertheless we show a trend for multi-view coronagraph observations improving CME arrival time forecast accuracy, and more events are needed to show this with 95\% statistical significance. For example, a future space weather mission at L5 or L4 as a second coronagraph viewpoint would reduce CME arrival time errors compared to a single L1 viewpoint. For example, \cite{akioka2005}, \cite{simunac2009}, \cite{gopalswamy2011}, \cite{strugarek2015}, \cite{vourlidas2015}, \cite{lavraud2016}, and \cite{weinzierl2016} all identify the potential benefit of an L5 mission. \cite{lavraud2016} proposes an L5 mission that would measure the coronal magnetic field using a polarization technique, in addition to white-light imaging. \cite{weinzierl2016} shows how an L5 mission equipped with a magnetic imager could improve predictions of solar wind, which is particularly important for the WSA-ENLIL+Cone model.

We also discussed various factors that should be taken into consideration when interpreting these verification results.  This includes uncertainties arising from the determination of CME input parameters from real-time coronagraph data, CME measurements from a variety of forecasters each with their own biases, the identification of ICME arrivals in-situ, glancing blow ICME arrivals, and multiple ICME arrivals. In future work, quality factors will be introduced for observed arrivals and identifying candidate CMEs. We will further quantify model performance by evaluating how well observed in-situ solar wind measurements compare to modeled values. We will also plan to assess CME arrivals at other locations, such as Mercury \citep{messenger} and Mars.

\begin{acknowledgements}
This work was completed while A.M.W. was an undergraduate student at American University, now a graduate student at University of Colorado Boulder. A.M.W. and M.L.M. thank Christine Verbeke and the CME Arrival Time and Impact Working Team (\href{https://ccmc.gsfc.nasa.gov/assessment/topics/helio-cme-arrival.php}{{\sf ccmc.gsfc.nasa.gov/assessment/topics/helio-cme-arrival.php}}) for useful discussions. M.L.M. acknowledges the support of NASA grant NNX15AB80G. L.K.J. acknowledges the support of NASA's STEREO project, NSF grants AGS 1321493 and 1259549.  D. Odstrcil acknowledges the support of NASA LWS-SC NNX13AI96G. The WSA
model was developed by N. Arge (AFRL), and the ENLIL model was
developed by D. Odstrcil (GMU). Estimated real-time planetary $K_{\rm P}$ indices are from NOAA as archived on CCMC's iSWA system (\href{https://ccmc.gsfc.nasa.gov/iswa/}{{\sf ccmc.gsfc.nasa.gov/iswa}}).
\end{acknowledgements}


\bibliography{enlilbib}


\end{document}